\documentclass[twocolumn,showpacs,preprintnumbers,amsmath,amssymb]{revtex4-1}
\usepackage{epsfig}
\usepackage{graphicx}
\usepackage{epsfig}
\usepackage{color}
\usepackage{rotating}
\usepackage{amssymb}
\usepackage{amsmath}
\usepackage{graphicx}
\usepackage{physics}

\def\beq{\begin{eqnarray}}
\def\eeq{\end{eqnarray}}

\newcommand{\be}{\begin{equation}}
\newcommand{\ee}{\end{equation}}
\newcommand{\bea}{\begin{eqnarray}}
\newcommand{\eea}{\end{eqnarray}}

\begin{document}

\title{The characteristic functions of quantum heat with baths at different temperatures}

\author{Erik Aurell}
\email{eaurell@kth.se}
\affiliation{
KTH -- Royal Institute of Technology,  AlbaNova University Center, SE-106 91~Stockholm, Sweden
}
\altaffiliation{
Depts. Computer Science and Applied Physics, Aalto University,  FIN-00076 Aalto, Finland
}

\begin{abstract}
This paper is about quantum heat defined as the change in energy of a bath during a process.
The presentation takes into account recent developments in classical strong-coupling thermodynamics,
and addresses a version of quantum heat which satisfies quantum-classical correspondence.
The characteristic function and the full counting statistics of quantum heat are shown to be formally similar.
The paper further shows that the method can be extended to more than one bath, \textit{e.g.} two baths at different temperatures,
which opens up the prospect of studying correlations and heat flow.
The paper extends earlier results on the expected quantum heat in the setting of one bath (Aurell \& Eichhorn 2015, Aurell 2017).
\end{abstract}


\maketitle

\section{Introduction}
\label{sec:Introduction}
The study of fluctuating work and heat in open quantum systems is an active interface
between non-equilibrium statistical physics and quantum information.
As statistical physics  
the starting point is
classical stochastic thermodynamics~\cite{sekimoto-book,Searles08-review,Jarzynski11-review,seifert12-review}
extended towards the quantum domain~\cite{JarzynskiWojcik2004,PhysRevE.73.046129,Esposito09-review,Campisi11-review,Chetrite2012,PhysRevLett.111.093602,PhysRevX.5.031038,PhysRevB.92.235440}.
In quantum information 
the term \textit{Quantum Thermodynamics}
has become standard~\cite{VinjanampathyAnders2016}; 
the approach is then full quantum descriptions of non-equilibrium
statistical physics of small systems often formulated as resource theories~\cite{PhysRevX.6.041017,PhysRevLett.115.210403}.

Fluctuation relations in classical stochastic thermodynamics
follow from ratios of path probabilities in forward and reverse processes.
This is so both for standard models of kinetic theory, where it has been
known for almost two decades~\cite{Maes99,ChG07,Gawedzki13},
as well as at strong coupling~\cite{Seifert2016,MillerAnders2017,Aurell2017b}.
In statistical physics a main problem of Quantum Thermodynamics 
is therefore that a quantum path is not an observable and so cannot be used as a 
conceptual building block of the theory. 
As quantum information the main problem is on the other hand that work and heat are not standard 
quantum operators in a Hilbert space, or in a space of density matrices.
Indeed, in Thermodynamics these quantities are not exact differentials and
are therefore properties of processes with a history, and not state functions.

One approach to Quantum Thermodynamics has its origin in the work of Feynman and Vernon~\cite{FeynmanVernon}.
An open quantum system is then explicitly modeled as a system of interest interacting
with a bath, and the bath variables are integrated out.
For a bath of harmonic oscillators initially in thermal equilibrium this
procedure can be carried out exactly and was used by Leggett and co-workers to investigate quantum Brownian motion, 
quantum tunneling and the spin-boson problem in three seminal papers in the 
1980ies~\cite{CaldeiraLeggett83a,CaldeiraLeggett83b,Leggett87}.
This theory, covered in several reviews and monographs \textit{cf.}~\cite{Grabert88,Weiss,BreuerPetruccione},
can be extended to also treat quantum heat as investigated recently by several 
groups~\cite{AurellEichhorn,Aurell2017,Carrega2015,Carrega2016,FunoQuan2017}.

The first goal of the present paper is to extend the results obtained 
for the expected value (first moment) in~\cite{AurellEichhorn} and~\cite{Aurell2017} 
to full generating functions of quantum heat. Analogous results have been obtained already in~\cite{Carrega2016}
and~\cite{FunoQuan2017}, up to differences in the physical set-up which
will be discussed in the following.
The second goal is to show that all the procedures in the paper can be
carried out for a system interacting with two or more baths at different temperatures.
As far as I am aware this observation is new.
The Feynman-Vernon approach thus open a way to investigate the generating functions
of quantum heat in a genuine non-equilibrium setting of heat flow between different reservoirs.       
As an example I derive an expression for quantum thermal power, the expected bath energy change per unit time in the long-time limit.

The organization of the paper is as follows.
%Section~\ref{sec:QFT-QT} gives an overview on quantum fluctutation theorems
%and quantum thermodynamics.
Section~\ref{sec:survey} states the problem studied, and the calculation for one bath oscillator 
is described in Sections~\ref{sec:FV}. Related technical details are given in Appendix~\ref{app:inversion}.
The setting of two or more baths at different temperatures 
and quantum thermal power are discussed in Section~\ref{sec:combining}.
Various conceptual and technical problem related to quantum heat 
are discussed in Section~\ref{sec:related}.
Section~\ref{sec:Discussion} sums up the paper and gives additional remarks. 

\section{Generating functions of quantum heat}
\label{sec:survey}
This Section builds extensively on~\cite{AurellEichhorn} and \cite{Aurell2017}. 
The setting is that of one quantum system (``the system'') linearly coupled to a large number of harmonic
oscillators `(``the baths''). The total Hamiltonian of the bath and the system is
\begin{equation}
H_{TOT} = H_S + \sum_{k} \left( H_B^{(k)} + H_I^{(k)} + H_C^{(k)}\right)
\label{eq:H_TOT}
\end{equation}
where 
\begin{equation}
H_S=\frac{P^2}{2M}+V(X,t)
\label{eq:H_S}
\end{equation}
is the system Hamiltonian (typically explicitly time dependent)
and $k$ labels the baths. When the distinction between the baths is not necessary the index $k$ will be dropped.
The bath Hamiltonians (for each bath) are thus
\begin{equation}
H_B = \sum_b \frac{p_b^2}{2m_b}+\frac{1}{2}m_b\omega_b^2 q_b^2.
\label{eq:H_B}
\end{equation}
where the mass and spring constant of oscillator $b$ are $m_b$ and $m_b\omega_b^2$; 
$\omega_b$ is the natural frequency in units $\hbox{rad}/s$, and $q_b$ and $p_b$
are respectively the oscillator's coordinate and momentum.
The interaction Hamiltonian is
\begin{equation}
H_I(t) = - \sum_b C_b(t)q_b X 
\label{eq:H_I}
\end{equation}
where the $C_b(t)$ are functions of time which start out at zero, rise up to a constant value
at the beginning of a process, and then go back to zero before the end of the process. 
The last term in the Hamiltonian is the Caldeira-Leggett counter-term (a correction to the system Hamiltonian)
$H_C(t) = \sum_b \frac{C_b^2(t)X^2}{2m_b\omega_b^2}$. The counter-term, the interaction term in (\ref{eq:H_I}) and the potential  
in (\ref{eq:H_B}) together form a sum of complete squares, $\sum_b \frac{1}{2}m_b\omega_b^2\left(q_b-\frac{C_b(t) X}{m_b\omega_b}\right)^2$.

Let now $\ket{i}$ and $\ket{f}$ be two states of the system, and let the 
system initially start in pure state $\dyad{i}$.
The oscillators in bath $k$ are initially in equilibrium at some inverse temperature $\beta_k$.
The total initial state of the system and the baths is thus
\begin{equation}
\rho_i^{TOT} = \rho_{B_1}^{eq}(\beta_1) \oplus  \rho_{B_2}^{eq}(\beta_2)  \oplus \cdots  \oplus \dyad{i}
\end{equation}
where $\rho_{B_k}^{eq}(\beta_k)$ is the Gibbs state of bath $k$ at inverse temperature $\beta_k$.

The goal of this paper is to compute the functionals (functionals of the system, functions of the parameters in the arguments)
\begin{widetext}
\begin{eqnarray}
\label{eq:charfunc}
F_{if}(\nu_1,\nu_2,\ldots) &=& \prod_k\hbox{Tr}_{B_k}\matrixel{f}{e^{i\left(\sum_k \nu_k H_{B_k}\right)}\left( U \rho_i^{TOT} U^{\dagger}\right) }{ f} \\
\label{eq:FCS}
G_{if}(\nu_1,\nu_2,\ldots) &=& \prod_k\hbox{Tr}_{B_k}\matrixel{f}{e^{i\left(\sum_k \nu_k H_{B_k}\right)}\left( U e^{-i\left(\sum_k \nu_k H_{B_k}\right)} \rho_i^{TOT} U^{\dagger}\right) }{ f}
\end{eqnarray}
\end{widetext}
where $\nu_1,\nu_2,\ldots$ are parameters that probe the energy of each bath.
The functional $F_{if}$ will be referred to as the characteristic function. With one bath and $i\nu_1=-\epsilon$
this quantity was computed in~\cite{AurellEichhorn} to first order in $\epsilon$.
The functional $G_{if}$ will be referred to as the full counting statistics (FCS). With one bath
this quantity was computed in~\cite{Aurell2017} to first order in $\nu_1$. 

As we will see in the following Sections $F_{if}$ and $G_{if}$ are structurally very similar, though not identical.
In fact, they only differ by the same kind of terms already found for their expected values in
respectively~\cite{AurellEichhorn} and~\cite{Aurell2017}.
$F_{if}$ is related to the generating function
of the energy of the baths in the final state, as can be seen as follows.
Consider one bath and the probability $p_B(E|i,f)$ of observing a bath energy $E$ at the final time, after the final measurement on the system.
The probability of observing \textit{one} state of the bath $\ket{\cal E}$, is
$\matrixel{\cal E}{\rho_B^{post}}{\cal E}$ where $\rho_B^{post}$ is the reduced density matrix of the bath,
after measuring the system and finding final state $\ket{f}$.
This quantity is (see \textit{e.g.} \cite{AurellEichhorn} before Eq~1)
\begin{equation}
\rho_B^{post}=\frac{1}{P_{if}} \matrixel{f}{\rho_{TOT}}{f}
\end{equation}
where $\rho_{TOT}$ is the total density operator of the system and the bath at the end of the process.
We then have
\begin{equation}
p_B(E|if) = \sum_{\cal E} \frac{1}{P_{if}} \mathbf{1}_{E({\cal E}),E} \matrixel{{\cal E},f}{\rho_B^{post}}{{\cal E},f}
\end{equation}
where $E({\cal E})$ is the energy of bath state $\ket{\cal E}$.
Integrating both sides one has
\begin{eqnarray}
\int e^{i\nu E} p_B(E|if) dE &=& \sum_{\cal E} \frac{1}{P_{if}} e^{i\nu E({\cal E})} \matrixel{{\cal E},f}{\rho_B^{post}}{{\cal E},f} \nonumber \\
                        &=& \frac{1}{P_{if}} F_{if}(\nu)
\end{eqnarray}
$F_{if}(\nu)$ is therefore up to the factor $P_{if}$ the generating function of the
bath energy in the final state.

$G_{if}(\nu)$ can on the other hand be related to a bath energy change in a two-measurement
protocol. If the initial state of the bath is both measured and recorded the initial state of 
the system and the bath is a pure state, $\ket{{\cal E}_i, i}$. 
If on the other hand the initial state of the bath is measured but not recorded 
the initial state is a statistical mixture where $\ket{{\cal E}_i, i}$ has Gibbs weight 
$Z_B^{-1}(\beta)\exp\left(-\beta E({\cal E}_i)\right)$. 
%Consider $p_B(\Delta E|if)$, the conditional probability of observing
%a change of bath energy $\Delta E$ conditioned on observing initial and final states $\ket{i}$ and $\ket{f}$ of the system. 
%We have
%\begin{equation}
%p_B(\Delta E|if) = \frac{p_B(\Delta E,f|i)}{p(f|i)}= \frac{1}{P_{if}} p_B(\Delta E,f|i)
%\end{equation}
Consider $p_B(\Delta E, f|{\cal E}_i, i)$, the conditional probability of observing
a final state $\ket{f}$ of the system and bath energy change
$\Delta E$, conditioned on total initial state 
$\ket{{\cal E}_i, i}$. 
Consider further $\overline{p}_B(\Delta E, f|i)=
\sum_{{\cal E}_i} p_B(\Delta E, f|{\cal E}_i, i)Z_B^{-1}(\beta)\exp\left(-\beta E({\cal E}_i)\right)$.
Similarly to above
\begin{eqnarray}
\overline{p}_B(\Delta E,f |i) &=& \sum_{{\cal E}_f,{\cal E}_i} Z_B^{-1}(\beta) e^{-\beta E({\cal E}_i)} \mathbf{1}_{E({\cal E}_f)-E({\cal E}_i),\Delta E} \nonumber \\
&&\quad \matrixel{{\cal E}_f,f}{\rho_{TOT}({\cal E}_i,i)}{{\cal E}_f,f}
\end{eqnarray}
where $\rho_{TOT}({\cal E}_i,i)$ is the total density operator of the system and the bath at the end 
of the process when the system and the bath started in the pure state $\ket{{\cal E}_i, i}$.
Resolving the delta function one can write 
\begin{equation}
\overline{p}_B(\Delta E,f|i)=\frac{1}{2\pi} \int e^{-i\nu\Delta E} G_{if}(\nu) d\nu 
\end{equation}
where 
\begin{eqnarray}
G_{if}(\nu) &=& \sum_{{\cal E}_f,{\cal E}_i} Z_B^{-1}(\beta)\exp^{-\beta E({\cal E}_i)} 
e^{i\nu\left(E({\cal E}_f)-E({\cal E}_i)\right)} \nonumber \\
&&\quad \matrixel{{\cal E}_f,f}{\rho_{TOT}({\cal E}_i,i)}{{\cal E}_f,f}
\end{eqnarray}
By linearity the Gibbs weight and the factor $e^{-i\nu E({\cal E}_i)}$
can be taken inside the the big unitary transformation defining
$\rho_{TOT}({\cal E}_i,i)$. The above is therefore alternatively
\begin{eqnarray}
\label{eq:FCS-MAIN}
G_{if}(\nu) &=& \hbox{Tr}_{B}\matrixel{f}{e^{i\nu H_{B}} U e^{-i\nu H_{B}} \rho_i^{TOT} U^{\dagger} }{ f}
\end{eqnarray}
which is the same expression as (\ref{eq:FCS}), for one bath.
$G_{if}(\nu)$, referred to a full counting statistics, is hence
the generating function of the energy change in the bath, averaged over an initial equilibrium distribution of the bath.

\section{The path integral expressions for the functionals $F_{if}$ and $G_{if}$}
\label{sec:FV}
Path integrals for harmonic oscillators can be done exactly since they are Gaussian.
As the initial state of the bath oscillators is factorized (they are independent)
and since they do not interact with one another, the path integral
of each bath oscillator can be done separately.

To keep the central message (and results) visible notional simplifications will be
introduced as needed.
The first such notation simplification, introduced by Feynman and Vernon~\cite{FeynmanVernon},
is to write $P_{if}$ for conditional probability of measuring the system in final state $\ket{f}$ given that
it was initially in $\ket{i}$.
Using the shorthand
\begin{equation}
\int_{if}\left(\cdots\right) = \int dX_i dY_i dX_f dY_f \psi_i(X_i)  \psi_i^*(Y_i) \psi_f^*(X_f)  \psi_f(Y_f) \left(\cdots\right) \nonumber
%\label{eq:if-def}
\end{equation}
where $\psi_i$ and $\psi_f$ are the wave functions of the states
$\ket{i}$ and $\ket{f}$. I write the transition probability as
\begin{equation}
P_{if} = \int_{if} {\cal D}X {\cal D}Y e^{\frac{i}{\hbar}S_S[X]-\frac{i}{\hbar}S_S[Y]+\frac{i}{\hbar}S_i[X,Y]-\frac{1}{\hbar}S_r[X,Y]} 
\label{P-if}
\end{equation}
where the two path integrals are over respectively the forward path $X(t)$ from $X_i$ to $X_f$
and the backward path $Y(t)$ from $Y_i$ to $Y_f$. These two path integrals emanate
from a representation of the total unitary $U$ and its inverse $U^{-1}$ in the time development of the
total density operator of the bath and the system 
$\hat{\rho}_{f}^{TOT} = U \hat{\rho}_{i}^{TOT}U^{-1}$, and then  
integrating out the bath variables.
The effects of the bath are thus captured by the two kernels
$S_i$ and $S_r$ in  (\ref{P-if}) which couple the forward and backward paths,
and which are referred to as the real and the imaginary part of the Feynman-Vernon action.
The contribution to $S_i$ and $S_r$ from \textit{one} oscillator with a time-dependent system-bath coupling is
\begin{widetext}
\begin{eqnarray}
\label{eq:S_i-b}
S_i^b &=& \int^t \int^s (X-Y)(X'+Y') \frac{C_b C_b'}{2m_b\omega_b} \sin\omega_b (s-s') ds' ds\\
\label{eq:S_r-b}
S_r^b &=& \int^t \int^s (X-Y)(X'-Y') \frac{C_b C_b'}{2m_b\omega_b} \coth\left(\frac{\omega\hbar\beta}{2}\right)\cos\omega_b (s-s') ds' ds 
\end{eqnarray}
\end{widetext}
where primed (unprimed) quantities refer to time $s'$ ($s$) and $\beta$ is the temperature of the bath to which this oscillator belongs.
The quantities in $S_i$ and $S_r$ in (\ref{P-if}) are the \textit{sums} of $S_i^b$ and $S_r^b$
from (\ref{eq:S_i-b}) and (\ref{eq:S_r-b}).
The expression for $S_r^b$ is symmetric in $s\leftrightarrow s'$ 
and this integral can therefore be extended over the whole square
$t_i \leq s,s'\leq t$.
The expression for $S_i^b$ is 
on the other hand not symmetric in $s\leftrightarrow s'$ 
so this integral has to be taken over the triangular domain
$t_i \leq s' \leq s\leq t$.

The path integral expressions for $F_{if}$ and $G_{if}$ can be written  
\begin{widetext}
\begin{eqnarray}
\label{eq:charfunc-FV}
F_{if} &=& \int_{if} {\cal D}X {\cal D}Y e^{\frac{i}{\hbar}S_S[X]-\frac{i}{\hbar}S_S[Y]}\prod_b {\cal F}_b^{(F)}(\nu) \\ 
\label{eq:FCS-FV}
G_{if} &=& \int_{if} {\cal D}X {\cal D}Y e^{\frac{i}{\hbar}S_S[X]-\frac{i}{\hbar}S_S[Y]}\prod_b {\cal F}_b^{(G)}(\nu)
\end{eqnarray}
\end{widetext}
where the products are over the bath oscillators $b$. The results of the corresponding integrations over
each bath oscillator are as in~\cite{AurellEichhorn}, Eq 13. I will write them as
\begin{widetext}
\begin{eqnarray}
\label{eq:F-factor-def}
{\cal F}_b^{(F)}(\nu) &=& \frac{\left(\frac{2\pi\hbar}{\omega m}\right)^2 e^{\frac{i}{2m\omega\hbar}\underline{u}\tilde{M_F}^{-1}\underline{u} +\frac{i}{\hbar}B}}
                         {Z(\beta)N(i\beta\hbar)|N(t)|^2 N(\hbar\nu) (\det\tilde{M_F})^{\frac{1}{2}}} \\
\label{eq:G-factor-def}
{\cal F}_b^{(G)}(\nu) &=& \frac{\left(\frac{2\pi\hbar}{\omega m}\right)^2 e^{\frac{i}{2m\omega\hbar}\underline{u}\tilde{M_G}^{-1}\underline{u} +\frac{i}{\hbar}B}}
                         {Z(\beta)N(i\beta\hbar-\hbar\nu)|N(t)|^2 N(\hbar\nu) (\det\tilde{M_G})^{\frac{1}{2}}} 
\end{eqnarray}
\end{widetext}
All quantities in above refer to one bath oscillator, the index $b$ understood in angular frequency $\omega$, mass $m$ etc.
The quantities $N(t)=\sqrt{\frac{2\pi i\hbar \sin (\omega t)}{m\omega}}$ in the denominators are the normalization factors of the harmonic oscillators
with the corresponding time-like arguments. The vector $\underline{u}$ 
and the function $B$ arise from the propagators of the harmonic oscillator
and are listed in Appendix~\ref{app:inversion}; they do not depend on $\nu$. 
The matrices $\tilde{M_F}$ and $\tilde{M_G}$ collect terms from the 
harmonic oscillator propagators, the initial equilibrium state of the bath, and 
$e^{i \nu H_B}$ and (for $G_{if}$) $e^{-i \nu H_B}$; they depend on
$\nu$ and are also listed in Appendix~\ref{app:inversion}. 

It was shown in~\cite{AurellEichhorn} that all the prefactors in (\ref{eq:F-factor-def}) combine to $\sinh\left(\frac{\omega\hbar\beta}{2}\right)
\sinh^{-1}\left(\frac{\omega\hbar(\beta-i\nu)}{2}\right)$. 
Using the same line of argument it is seen that all the prefactors in (\ref{eq:G-factor-def}) combine to give one.
The whole non-trivial part hence stems from $\underline{u}$, $B$ and $\tilde{M}$
and the results can be written as    
\begin{widetext}
\begin{eqnarray}
\label{eq:expr-MAIN-F}
\prod_b {\cal F}^{(F)}_b(\nu) &=& \left(\hbox{trivial}\right)\cdot e^{\frac{i}{\hbar} S_i -  \frac{1}{\hbar} S_r + {\cal J}^{(1)}(\nu) + {\cal J}^{(2)}(\nu) + {\cal J}^{(3)}(\nu)} \\
\label{eq:expr-MAIN-G}
\prod_b {\cal F}^{(G)}_b(\nu) &=& e^{\frac{i}{\hbar} S_i -  \frac{1}{\hbar} S_r  + {\cal J}^{(2)}(\nu) + {\cal J}^{(3)}(\nu)} 
\end{eqnarray}
where the Feynman-Vernon terms $S_i$ and $S_r$ are given in (\ref{eq:S_i-b}) and (\ref{eq:S_r-b}).
The three new functionals, which
are all symmetric in $s\leftrightarrow s'$, are
%\begin{widetext}
\begin{eqnarray}
\label{eq:J1-def-MAIN}
{\cal J}^{(1)} &=& \frac{i}{2m\omega\hbar}\int^t\int^t (XX'+YY')CC' \cos\omega(s-s') \left(\frac{y-z}{\Delta} -\frac{i}{2}\coth\frac{\omega\hbar\beta}{2}\right) \\
\label{eq:J2-def-MAIN}
{\cal J}^{(2)} &=& \frac{i}{2m\omega\hbar}\int^t\int^t (XY'-X'Y)CC' \sin\omega(s-s') \left(\frac{y'z'-yz}{\Delta} -\frac{1}{2}\right) \\
\label{eq:J3-def-MAIN}
{\cal J}^{(3)} &=& \frac{i}{2m\omega\hbar}\int^t\int^t (XY'+X'Y)CC' \cos\omega(s-s') \left(\frac{z'-y'}{\Delta} +\frac{i}{2}\coth\frac{\omega\hbar\beta}{2}\right)
\end{eqnarray}
\end{widetext}
where the auxiliary variables $z$, $z'$,  $y$, $y'$ and $\Delta$ are 
combinations of trigonometric and hyperbolic functions in $\nu$ and $\beta$ given
in Appendix~\ref{app:inversion}.
The definitions 
of $z$, $z'$ and $\Delta$ differ a bit between Case~F ($F_{if}$) and Case~G ($G_{if}$).
However, for both cases ${\cal J}^{(1)}$, ${\cal J}^{(2)}$ and ${\cal J}^{(3)}$ all
vanish at $\nu=0$.
For Case G ($G_{if}$) the functional ${\cal J}^{(1)}$ as defined by (\ref{eq:J1-def-MAIN})
is identically zero, so that the only remaining pieces are ${\cal J}^{(2)}$ and ${\cal J}^{(3)}$,
in agreement with the expression given in (\ref{eq:expr-MAIN-G}).

\section{Quantum heat flows and quantum thermal power}
\label{sec:combining}
The first point of this section is that if a system
interacts with two or more harmonic oscillator heat baths that do not interact directly
with one another, the corresponding Feynman-Vernon actions simply add.
The reason is the same as used to derive the Feynman-Vernon action 
from one bath by adding the contributions from each oscillator separately.

The second point is that the same property holds for 
the generating functions 
$F_{if}(\nu_1,\nu_2,\ldots)$
and $G_{if}(\nu_1,\nu_2,\ldots)$
introduced in (\ref{eq:charfunc}) and (\ref{eq:FCS}).
For the rest of this section I will assume that the system interacts 
with two baths, a cold (``left'') bath at inverse temperature $\beta_L$,
and a hot (``right'') bath at inverse temperature $\beta_R$ ($\beta_R<\beta_L$).
It is natural to expect that if the system has some structure
all parts of the system do not interact with two baths in the same
way. Previously it was not necessary to make this distinction,
but here it is convenient to think of one system coordinate $X_L$
which interacts linearly with the cold bath, and one system coordinate $X_R$
which interacts linearly with the hot bath, and all the other system coordinates grouped into $X_C$  
Classically one then expects heat to flow from right to left 
from the hot bath to $X_R$, then from $X_R$ through the system to $X_L$, and then from $X_L$ out into the cold bath.
A similar setting was recently considered in~\cite{KarimiPekola2017}.

The main interest should be in the long-time limit of averages and fluctuations
of the energy changes in the baths,
and it reasonable to assume that these will not depend much on the initial or the final state of the system.
Let the average of the energy change(s) in the bath(s) over the final state of the system be
\begin{equation}
\overline{p}_B(\Delta E|i)=\sum_f \overline{p}_B(\Delta E, f|i) 
\end{equation}
and let $G_i(\nu_1,\nu_2)$ be the corresponding generating functions.
We have 
\begin{equation}
G_i(\nu_1,\nu_2) =\sum_f G_{if}(\nu_1,\nu_2)  
\end{equation}
The starting point is then
\begin{widetext}
\begin{eqnarray}
\label{eq:FCS-2}
G_i(\nu_1,\nu_2) &=& \hbox{Tr}_{B_L,B_R,S}\left[ e^{i\left(\nu_1 H_{B_L}+\nu_2 H_{B_R}\right)}\left( U e^{-i\left(\nu_1 H_{B_L}+\nu_2 H_{B_R}\right)} \rho_i^{TOT} U^{\dagger}\right)\right] \nonumber \\
&=& \int_i {\cal D}X {\cal D}Y e^{\frac{i}{\hbar}S_S[X]-\frac{i}{\hbar}S_S[Y]}\prod_{b\in L} {\cal F}_b^{(G)}(\nu_1) \prod_{b\in R} {\cal F}_b^{(G)}(\nu_2) 
\, \delta(X^f-Y^f)
\end{eqnarray}
where the trace is over the system and both baths and $\delta(X^f-Y^f)$ is the path integral rendering of $\hbox{Tr}_{S}$.
The subscript of the integral $\int_i$ indicates the remaining dependence on the initial state of the system. The two products in
(\ref{eq:FCS-2}) are
\begin{eqnarray}
\label{eq:expr-MAIN-G-C}
\prod_{b\in L} {\cal F}^{(G)}_b(\nu_1) &=& e^{\frac{i}{\hbar} S_i[X_L,Y_L] -  \frac{1}{\hbar} S_r[X_L,Y_L]  + {\cal J}^{(2)}(\nu)[X_L,Y_L] + {\cal J}^{(3)}(\nu)[X_L,Y_L]} \\
\label{eq:expr-MAIN-G-H}
\prod_{b\in R} {\cal F}^{(G)}_b(\nu_2) &=& e^{\frac{i}{\hbar} S_i[X_R,Y_R] -  \frac{1}{\hbar} S_r[X_R,Y_R]  + {\cal J}^{(2)}(\nu)[X_R,Y_R] + {\cal J}^{(3)}(\nu)[X_R,Y_R]} 
\end{eqnarray}
\end{widetext}
and the functionals are given as sums of the terms in (\ref{eq:S_i-b}), (\ref{eq:S_r-b}) and
(\ref{eq:J1-def-MAIN}-\ref{eq:J3-def-MAIN}).
Averages, correlations and cross-correlations of the energy changes in the two baths can be evaluated
as derivatives of $G_i$ with respect to $\nu_1$ and $\nu_2$.

Quantum thermal machine are quantum systems that transform heat to useful work.
Quantum analogues of Carnot, Otto and Diesel engines as well as
quantum refrigerators have been proposed and partly experimentally realized~\cite{Pekola2015,1367-2630-17-3-035012},
and were reviewed in~\cite{VinjanampathyAnders2016}.
The amount of heat flowing through the working fluid of a quantum thermal machine
per unit of time limits how much work the machine can do per unit of time \textit{i.e.} the power. 
The simplest quantity that can
be considered by the above analysis is quantum thermal power
defined 
as the energy change in the one of the baths per unit time, in the limit when
the process is in steady state. 

To be concrete, let the bath be the cold bath.
One can expect the dependence on the initial state $\ket{i}$
of the system to drop out, and 
one can write 
\begin{widetext}
\begin{eqnarray}
\dot{Q} = \lim_{t_f-t_i\to\infty} \frac{\partial_{i\nu_1}G_i|_{\mathbf{\nu}=0}}{t_f-t_i} &=& \int_{-\infty}^0h^{(2)}(0,s)\left<\left(X_L(0)Y_L(s)-X_L(s)Y_L(0)\right)\delta(X_L(0)-Y_L(0))\right> ds \nonumber \\
     &&\, + \int_{-\infty}^0h^{(3)}(0,s)\left<\left(X_L(0)Y_L(s)+X_L(s)Y_L(0)\right)\delta(X_L(0)-Y_L(0))\right> ds
\label{eq:quantum-thermal-power}
\end{eqnarray}
where the two kernels are the 
terms linear in $\nu_1$ of ${\cal J}^{(2)}$ and
${\cal J}^{(3)}$ as given in (\ref{eq:J2-der-G}) and (\ref{eq:J3-der-G}).
\begin{eqnarray}
\label{eq:h2-definition}
h^{(2)}(s,s')&=& i\sum_b\frac{C_b^2}{2m_b}\coth(\frac{\beta\hbar\omega_b}{2})\sin\omega_b(s-s') \\
\label{eq:h3-definition}
h^{(3)}(s,s')&=& \sum_b\frac{C_b^2}{2m_b}\cos\omega_b(s-s')
\end{eqnarray}
\end{widetext}
Up to a factor $\beta$ the above is the same as Eq~16 in~\cite{AurellEichhorn}).
As we consider a long-time limit the time-dependence of $C_b$ in the equivalent expressions 
in (\ref{eq:J2-der-G}) and (\ref{eq:J3-der-G}) can be ignored.
The expectation values in (\ref{eq:quantum-thermal-power}) are over a 
steady-state reduced description of ``left'' part of 
the system only.

An interesting special case is when the bath is Ohmic 
and the temperature (in this case, of the cold bath) is sufficiently high (Caldeira-Leggett limit).
For a system with continuous state space this yields the 
classical limit of stochastic thermodynamics (Kramers-Langevin equation). 
The two kernels can then be approximated as $h^{(2)}(s,s')\approx -\frac{2i}{\beta_L\hbar}\eta \frac{d\delta(s-s')}{d(s-s')}$
and $h^{(3)}(s,s')\approx - \eta \frac{d^2\delta(s-s')}{d(s-s')^2}$~\cite{Aurell2017}
where $\eta$ is the friction coefficient. Integration by parts turns (\ref{eq:quantum-thermal-power})
into
\begin{eqnarray}
  \dot{Q} &=& \int dX dY \delta(X-Y) \frac{i\eta}{\hbar\beta} \left<\dot{X}Y-X\dot{Y}\right> \overline{\rho}_L(X,Y) \nonumber \\
             && + \int dX dY \delta(X-Y) 2\eta \left<\dot{X}\dot{Y}\right> \overline{\rho}_L(X,Y) 
\label{eq:I1-I2-I3-average-4}
\end{eqnarray}
where $\overline{\rho}_L(X,Y)$ is the stationary reduced density matrix of the left part of the system in the coordinate representation.

When the system has continuous state space the averages in (\ref{eq:I1-I2-I3-average-4}) and over the system development 
in the Caldeira-Leggett limit
and were evaluated in~\cite{AurellEichhorn} Sections 6.2 and 6.4. The result can be expressed by two operators
\begin{eqnarray}
  \hat{O}_1 &=& -2\frac{\eta}{M}\frac{\partial}{\partial(X-Y)} (X-Y) \\
    \hat{O}_2 &=& -2\eta\hbar^2\frac{\eta}{M^2}\frac{\partial^2}{\partial(X-Y)^2}  
\end{eqnarray}
and
\begin{eqnarray}
  \dot{Q} &=& \hbox{Tr}\left[  \hat{O}_1 \overline{\rho}_S \right] + \hbox{Tr}\left[  \hat{O}_2 \overline{\rho}_S \right] 
\label{eq:ThermalPowerOperators}
\end{eqnarray}
The Caldeira-Leggett limit is essentially a classical limit because the Wigner transform of the density matrix obeys
classical Fokker-Planck equation. The above is hence just a quantum way of writing the heat per unit time
as in classical stochastic thermodynamics.

For a quantum system with discrete state space (some number of qubits), one would have to
go back to (\ref{eq:I1-I2-I3-average-4}) for an Ohmic bath.
If the qubits interact only through one spin component,
say through $\hat{\sigma}_z$, then the dynamics of the system
has a path integral representation 
first introduced by Leggett and co-workers for the spin-boson problem~\cite{Leggett87}.
This approach (with or without a bath) has been developed further in the statistical physics
community to model quantum annealing protocols~\cite{Jorg2008,Bapst2013}.
If on the other hand the qubits interact in a more general manner,
as they would theoretically have to in general-purpose quantum computational device,
then a more involved path integral representation would have to be used~\cite{Klauder79,Stone89,AurellQC}.
For a quantum system interacting with baths that are not Ohmic
(non-Markovian state evolution) one would have to go back to (\ref{eq:quantum-thermal-power}).

\section{Conceptual and technical problems of quantum heat}
\label{sec:related}
In this paper I have defined quantum heat as the changes of bath energy 
when the baths are initially in thermal equilibrium and the system-bath
interaction vanishes at the beginning and the end of a process.
This translates to the quantum domain the strong-coupling classical definition
of heat introduced in~\cite{Aurell2017}.
It could have been assumed that it would be simpler to take
the system-bath interaction constant and to somehow estimate quantum heat from
the bath Hamiltonian and interaction Hamiltonian at the initial and final
time. The section discusses why such an approach is not straight-forward.

A first indication of a problem is that if one would simply take 
quantum heat as change of bath energy and the system-bath interaction 
constant in time there appears in the classical limit (Caldeira-Leggett model)
boundary contributions, discussed at length in~\cite{AurellEichhorn}.
In such an approach there is hence not a complete quantum-classical correspondence
for heat even on the level of expectation values.

A second indication comes from the recent development of strong-coupling stochastic 
thermodynamics~\cite{Seifert2016,Jarzynski2017}.
Even classically it is only when the system-bath interaction energy is negligible
that one can at the same time take it constant and define heat as change in bath energy.
If the system-bath interaction energy is comparable to changes in system energy
or bath energy then a fraction should be counted as heat, and a fraction as change
of system energy. 
Certain choices of these fractions, where also part of the bath
energy is counted as internal energy of the system, have been found to
be consistent, albeit at the price that the resulting heat 
has to determined by thermodynamic integration.
The latter has led to a vigorous polemic~\cite{TalknerHanggi2016}
which I have recently discussed elsewhere~\cite{Aurell2017b}.
The main problem in the present (quantum) context is that even when the proposals in~\cite{Seifert2016,Jarzynski2017}
can be accepted classically, they lead to quite involved definitions in the quantum domain, compare Eq. 28 in~\cite{Seifert2016} and the discussion in~\cite{MillerAnders2018}.
The alternative procedure of a time-dependent system-bath interaction
avoids this problem on both the classical and the quantum level, at the price of a new term in the work~\cite{Aurell2017}.

It can be concluded that quantum heat is in some sense always a strong-coupling phenomenon,
and problems with various na\"{i}ve versions of heat in open quantum systems 
indeed central issues in Quantum Thermodynamics.
The earliest indication may have been~\cite{daCosta2000} where it was shown that
the quantum dynamics of a single oscillator coupled to a heat bath of harmonic oscillators
depends sensitively on the exact initial conditions, and in particular if the bath is brought into
contact immediately before or immediately after the system is measured initially.
A second problem was identified in~\cite{Hanggi2006,Hanggi2008} where it was shown
that different definitions of specific heat of a quantum particle interacting 
strongly with a bosonic heat bath yield different results.
A further step was taken in~\cite{PhysRevB.92.235440}
where it was shown that for a system interacting with a fermonic bath
one cannot consistently include any fixed non-zero fraction of the system-bath interaction
in the heat. This excludes, for instance, the choice of including all system-bath interaction, 
as proposed in~\cite{AllahverdyanNieuwenhuizen2001}.
On the other hand, including no part of the system-bath interaction
in the heat means to treat it as if weakly coupled, which brings 
the issue that Third Law is no longer satisfied~\cite{PhysRevB.92.235440}.
Note that the choice in~\cite{Seifert2016,Jarzynski2017} includes
(classically) a definite but not a fixed fraction of the system-bath interaction energy in the heat.

Earlier work technically most similar to the determination 
of the generating function of heat in this paper are~\cite{Carrega2015},~\cite{Carrega2016} and \cite{FunoQuan2017}.
Those papers mostly appeared before the recent developments of (classical) strong-coupling
stochastic thermodynamics, and were hence developed independently of that context.
In~\cite{Carrega2016} was computed a generating function formed from 
inserting the operators $e^{i\nu H_B+i\nu\lambda H_{SB}}$
at the final time and $e^{-i\nu H_B}$ at the initial time; $\lambda$ is here an additional parameter.
The interaction is with one bath at inverse temperature $\beta$, and leads to expressions of a 
similar structure to the ones given above for $G_{if}$.
The expectation value formed 
by differentiating this generating function with respect to $i\nu$ at $\nu=0$ 
is $<H_B+\lambda H_{SB}>_f - <H_B>_i$ 
where $<\cdots>_i$ means averaging with respect to the
initial state of the bath (independent of the system) 
and  $<\cdots>_f$ means averaging with respect to the final state of the system and the bath.
By the discussion above this choice
does not correspond classically to any of the proposals currently considered viable
for strong-coupling heat in stochastic thermodynamics, the only exception being
$\lambda=0$ and weak coupling. 
One may note that the boundary contribution in the classical limit from~\cite{AurellEichhorn}  
does not appear in the formulation in~\cite{Carrega2016}, 
at least not at the initial time, due to an additional assumption that the initial state of the system is diagonal.
In~\cite{FunoQuan2017} (v1 as available on arXiv, Supplementary Material Section III) a similar calculation is carried
out from the operators
$e^{-i\nu\left( H_B+H_{SB}\right)}$ at the final time 
and $e^{-i\nu\left( H_B+H_{SB}\right)}$ at the initial time,
with the bath and system initially in joint equilibrium.
This leads again to expressions of a similar structure to the ones given above for $G_{if}$.
The choice of an initial joint equilibrium state of the system and the bath
is the same as in~\cite{Hanggi2006,Hanggi2008}.
Classically it can be seen as special case of of the choice in~\cite{Seifert2016,Jarzynski2017} 
when the initial state of the system is an equilibrium at mean force, for a comparison see~\cite{Aurell2017b}.
The corresponding expectation value 
is $<H_B+H_{SB}>_f - <H_++H_{SB}>_i$ where both averages are over the system and the bath
implies the same definition of quantum heat as used in~\cite{AllahverdyanNieuwenhuizen2001}.

\section{Discussion}
\label{sec:Discussion}
I have in this paper computed the generating functions of
the distributions of the final energy in a bosonic bath (or baths) and the change
of bath energy as functionals of a system interacting with the bath (or baths).
From a technical point of view analogous results
were obtained~\cite{Carrega2016} and~\cite{FunoQuan2017}
but in slightly different settings which do fit the recently developed 
(classical) stochastic thermodynamics at strong coupling~\cite{Seifert2016,Jarzynski2017}.
The generating functions computed here directly generalize earlier 
results on the expected value (first moment) obtained in~\cite{AurellEichhorn} and~\cite{Aurell2017}. 

The most remarkable analytic properties of all these result are quite explicit formulae for
the generating functions. These are quadratic functionals
of the forward and backward paths in the Feynman-Vernon formalism,
with kernels of a similar type as for the real and imaginary parts of the Feynman-Vernon action
\textit{i.e.} combinations of trigonometric and hyperbolic functions of
time differences, bath oscillator frequencies, bath temperatures and the generating function parameters.
The first and second derivatives of the generating functions at the origin
have been evaluated (in Appendix~\ref{app:inversion}) and determine
the expected value, variances and cross-correlations of bath energy changes.

I have also in this paper pointed out that the extended Feynman-Vernon approach works equally well
for systems interacting with more than one bath at different temperatures.
It is therefore a principled way to define and estimate
non-equilibrium quantum heat flows. As an example I have
derived the quantum thermal power of a system connected to an Ohmic heat bath
and showed that it agrees with average power in the classical limit. 

The real potential advantage of the approach developed
here would be it could also in practice be applied to systems with 
discrete states.
Superconducting qubits is the
currently favored platform for quantum computing
and~\cite{Wendin2017}; every computing element or
``qubit'' is then in fact a degree of freedom of a large (mesoscopic) object
at very low temperature~\cite{Devoret1995}.
Understanding heat flow and other thermal properties
of such objects is an active area of research~\cite{Pekola2015}
where a general theoretical frame-work so far has been lacking.
In this context it is noteworthy that generating functions of the type
considered here (with one bath) already were applied to  
the spin-boson problem in~\cite{Carrega2016}.

Finally, a thermal bath consisting of harmonic oscillators is a model of delocalized
environmental modes such as phonons. The main degrees of freedom in a real material at very low temperature, such as
defects and nuclear spins, are on the other hand likely to be localized, and may be more 
accurately described as a spin bath~\cite{ProkofevStamp2000}. 
Path integral representations of systems interacting with such spin baths
were developed quite some time ago~\cite{Chen1987},
and could potentially be extended to also describe heat flows between such baths.

\section*{Acknowledgments}
I thank Ralf Eichhorn, Ken Funo, Yuri Galperin, Bayan Karimi, Jukka Pekola and  
Haitao Quan for many discussions and constructive remarks.
This research was supported by the Academy of Finland through its Center of Excellence COIN
and by the Chinese Academy of Sciences CAS
President’s International Fellowship Initiative (PIFI)
grant No. 2016VMA002.

\appendix

\section{Inverting the matrices and determining the functionals}
\label{app:inversion}
\begin{widetext}
This appendix contains the derivation of equations
(\ref{eq:expr-MAIN-F}-\ref{eq:J3-def-MAIN})
in the main text, and then the closed-form expressions of the
kernels ${\cal J}^{(1)}$, ${\cal J}^{(2)}$ and ${\cal J}^{(3)}$ for case F and case G.
Conventions are as in~\cite{AurellEichhorn} except that a factor $i$ 
for convenience has been included in the definitions of the matrices $\tilde{M}$.
The vector $\underline{u}$, appropriate for when the coupling coefficient depends on time, is
\begin{equation}
\label{eq:u-def}
\underline{u} = \left(\begin{array}{lcl} u &=& \frac{1}{\sin\omega t}\int^t \sin\omega(t-s) [C(s)X(s)]ds \\
                                         v &=& \frac{1}{\sin\omega t}\int^t \sin\omega(t-s) [-C(s)Y(s)] ds \\
                                         u'&=& \frac{1}{\sin\omega t}\int^t \sin\omega s [C(s)X(s)]ds \\
                                         v'&=& \frac{1}{\sin\omega t}\int^t \sin\omega s [-C(s)Y(s)] ds 
                       \end{array}\right)
\end{equation}
The function $B$ is similarly 
\begin{eqnarray}
B &=& - \frac{1}{m\omega\sin (\omega t)}\int^t \int^s  \sin\omega(t-s) \sin\omega s' C X C' X' ds' ds \nonumber \\
&&+ \frac{1}{m\omega\sin (\omega t)}\int^t \int^s  \sin\omega(t-s) \sin\omega s' C Y C' Y' ds' ds
\label{eq:B-def}
\end{eqnarray}
where the primed (unprimed) quantities refer to time $s'$ ($s$).

The matrix $\tilde{M}$ can be written in the same way for the two cases
by introducing auxiliary variables:
\begin{equation}
\label{eq:M-def}
\tilde{M} = \left(\begin{array}{llll} -x-z & z' & x' & 0 \\
                                      z' & x-z & 0 & x' \\
                                      x' & 0 & -x+y & -y' \\
                                      -x-z & z' & -y' & x+y 
                       \end{array}\right)
\end{equation}
These auxiliary variables are the same as in equations (\ref{eq:J1-def-MAIN}-\ref{eq:J3-def-MAIN})
in the main text, and are defined as follows:
\par\noindent
\textbf{Case F:} In this case $x=\cot(\omega t)$, $x'=\sin^{-1}(\omega t)$,
$y=\cot(\omega\hbar \nu)$, $y'=\sin^{-1}(\omega\hbar\nu)$,
$z=\cot(-i\omega\hbar\beta)=i \coth(\omega\hbar\beta)$ and $z'=\sin^{-1}(-i\omega\hbar\beta)=i\sinh^{-1}(\omega\hbar\beta)$.
\par\noindent
\textbf{Case G:} In this case $x$, $x'$, $y$ and $y'$ are the same as in case F
while $z=\cot(\omega\hbar(\nu-i\beta))$ and $z'=\sin^{-1}(\omega\hbar(\nu-i\beta))$.
\par\noindent
\textbf{Algebraic relations:} In both cases the auxiliary variables satisfy obvious relations, namely
\begin{equation}
\label{eq:alg-relations}
z'^2-z^2 = y'^2-y^2 = x'^2-x^2 = 1
\end{equation}
Using (\ref{eq:alg-relations}) repeatedly it is straight-forward to determine the matrix inverse as
\begin{equation}
\label{eq:M-inv}
\tilde{M}^{-1} = \frac{1}{\Delta} \left(\begin{array}{llll} y-z & y'-z' & D & -B \\
                                                           y'-z' & y-z & -C & A \\
                                                           D & -C & y-z & y'-z' \\
                                                           -B & A & y'-z' & y-z 
                       \end{array}\right)
\end{equation}
where new auxiliary variables are
\begin{eqnarray}
A &=& \frac{1}{x'}\left(1+x(y-z) +yz - y'z'\right) \\
B &=& \frac{1}{x'}\left(x(z'-y') +y'z - yz'\right) \\
C &=& \frac{1}{x'}\left(x(z'-y') -y'z + yz'\right) \\
D &=& \frac{1}{x'}\left(-1+x(y-z) - yz + y'z'\right) 
\end{eqnarray}
and
\begin{equation}
\Delta = AD-BC = 2\left(y'z'-yz-1\right).
\end{equation}
The combination that enters the exponent in (\ref{eq:F-factor-def})
and (\ref{eq:G-factor-def}) in the main text is thus
\begin{eqnarray}
\underline{u}\tilde{M}^{-1}\underline{u} &=& \frac{1}{\Delta} \big( (y-z)\left(u^2+v^2+u'^2+v'^2\right) \nonumber \\
&& + (y'-z')\left(2uv+2u'v'\right) +2Duu' \nonumber \\
&& - 2Buv'  - 2Cvu' + 2Avv'\big)   
\end{eqnarray}
where $u$, $v$, $u'$ and $v'$ are given in (\ref{eq:u-def}).
Combining this with the term $\frac{i}{\hbar}B$ and using trigonometric identities
the whole expression reduces to
\begin{equation}
\label{eq:expr}
\hbox{Expr.} = \frac{i}{\hbar} S_i -  \frac{1}{\hbar} S_r + {\cal J}^{(1)} + {\cal J}^{(2)} + {\cal J}^{(3)}
\end{equation}
where the Feynman-Vernon terms $S_i$ and $S_r$ are given in (\ref{eq:S_i-b}) and (\ref{eq:S_r-b}).
Restating for convenience here equations (\ref{eq:J1-def-MAIN}-\ref{eq:J3-def-MAIN}) in the main text they are 

\begin{eqnarray}
\label{eq:J1-def}
{\cal J}^{(1)} &=& \frac{i}{2m\omega\hbar}\int^t\int^t (XX'+YY')CC' \cos\omega(s-s') \left(\frac{y-z}{\Delta} -\frac{i}{2}\coth\frac{\omega\hbar\beta}{2}\right) \\
\label{eq:J2-def}
{\cal J}^{(2)} &=& \frac{i}{2m\omega\hbar}\int^t\int^t (XY'-X'Y)CC' \sin\omega(s-s') \left(\frac{y'z'-yz}{\Delta} -\frac{1}{2}\right) \\
\label{eq:J3-def}
{\cal J}^{(3)} &=& \frac{i}{2m\omega\hbar}\int^t\int^t (XY'+X'Y)CC' \cos\omega(s-s') \left(\frac{z'-y'}{\Delta} +\frac{i}{2}\coth\frac{\omega\hbar\beta}{2}\right)
\end{eqnarray}
The only difference in the expressions for cases \textbf{F} and \textbf{G} are the different interpretations
of the auxiliary variables $z$, $z'$ and $\Delta$. 
We now proceed to simplify the coefficients in the kernels in the two cases.

\subsection{Case F}
We here have 
\begin{eqnarray}
\Delta = 2(z'y'-yz-1) &=& 2i\sinh^{-1}(\omega\hbar\beta)\sin^{-1}(\omega\hbar \nu)\left(1-\cos(\omega\hbar \nu)\cosh(\omega\hbar\beta)+i \sin(\omega\hbar \nu) \sinh(\omega\hbar\beta)\right) \nonumber \\
&=& 2i\sinh^{-1}(\omega\hbar\beta)\sin^{-1}(\omega\hbar \nu)\left(1-\cos(\omega\hbar(\nu+i\beta))\right) 
\end{eqnarray}
and the expressions simplify to 
\begin{eqnarray}
\frac{y-z}{\Delta}     &=& \frac{1}{2}\frac{\cos(\omega\hbar \nu)\sin(\omega\hbar(-i\beta)) - \sin(\omega\hbar \nu)\cos(\omega\hbar(-i\beta))}
                                           {1-\cos(\omega\hbar(\nu+i\beta))} = \frac{1}{2} \cot\left(\frac{\omega\hbar(\nu+i\beta)}{2}\right) \nonumber \\
\frac{y'z'-yz}{\Delta} &=& \frac{1}{2} \frac{1- \cos(\omega\hbar \nu)\cos(\omega\hbar(-i\beta))}
                                               {1-\cos(\omega\hbar(\nu+i\beta))}=\frac{1}{2} 
+\frac{i}{4}\frac{\sin(\omega\hbar \nu)\sinh(\omega\hbar\beta)}{\sin^2(\frac{\omega\hbar(\nu+i\beta)}{2})}
\nonumber \\
\frac{z'-y'}{\Delta}   &=& \frac{1}{2}\frac{\sin(\omega\hbar \nu)+i\sinh(\omega\hbar\beta)}
                                            {1-\cos(\omega\hbar(\nu+i\beta))}  = +\frac{1}{4}
\frac{\sin(\omega\hbar \nu)+i\sinh(\omega\hbar\beta)}
                                            {\sin^2(\frac{\omega\hbar(\nu+i\beta)}{2})}
\nonumber
\end{eqnarray}
All three functionals ${\cal J}^{(1)}$, ${\cal J}^{(2)}$ and
${\cal J}^{(2)}$ vanish at $\nu=0$ as follows by insertion in above.
Their derivatives at $\nu=0$ are
\begin{eqnarray}
\label{eq:J1-der-F}
\frac{\partial {\cal J}^{(1)}}{\partial\nu}|_{\nu=0} &=& -\frac{i}{8m}\int^t\int^t (XX'+YY')CC' \cos\omega(s-s') \frac{1}{\sinh^2(\frac{\omega\hbar\beta}{2})} \\
\label{eq:J2-der-F}
\frac{\partial {\cal J}^{(2)}}{\partial\nu}|_{\nu=0} &=& -\frac{1}{4m}\int^t\int^t (XY'-X'Y)CC' \sin\omega(s-s') \coth(\frac{\omega\hbar\beta}{2}) \\
\label{eq:J3-der-F}
\frac{\partial {\cal J}^{(3)}}{\partial\nu}|_{\nu=0} &=& \frac{i}{8m}\int^t\int^t (XY'+X'Y)CC' \cos\omega(s-s')
\left(\frac{1}{\sinh^2(\frac{\omega\hbar\beta}{2})}+ 2  \right)  
\end{eqnarray}
The derivative of the second term $(y'z'-yz)/\Delta$ gives the term called ${\cal I}^{(2)}$ in~\cite{AurellEichhorn} and~\cite{Aurell2017}. 
while the derivatives of $(y-z)/\Delta$ and $(z'-y')/\Delta$ 
combine to give the terms called ${\cal I}^{(1)}$ and ${\cal I}^{(3)}$ in~\cite{AurellEichhorn},
with the idenification $\nu=i\epsilon$ and a factor two from the definition of the double integrals.

\subsection{Case G}
We here have 
\begin{eqnarray}
\Delta = 2(z'y'-yz-1) &=& 2\sin^{-1}(\omega\hbar(\nu-i\beta))\sin^{-1}(\omega\hbar \nu)\left(1-\cos(\omega\hbar \nu)\cos(\omega\hbar(\nu-i\beta))- \sin(\omega\hbar \nu) \sin(\omega\hbar(\nu-i\beta))\right) \nonumber \\
&=& 2\sin^{-1}(\omega\hbar(\nu-i\beta))\sin^{-1}(\omega\hbar \nu)\left(1-\cosh(\omega\hbar\beta)\right) 
\end{eqnarray}
and the expressions simplify a bit further to
\begin{eqnarray}
\frac{y-z}{\Delta}     &=& \frac{1}{2}\frac{\cos(\omega\hbar \nu)\sin(\omega\hbar(\nu-i\beta)) - \sin(\omega\hbar \nu)\cos(\omega\hbar(\nu-i\beta))}
                                           {1-\cosh(\omega\hbar\beta)} = \frac{i}{2} \coth\left(\frac{\omega\hbar\beta}{2}\right) \nonumber \\
\frac{y'z'-yz}{\Delta} &=& \frac{1}{2} \frac{1- \cos(\omega\hbar \nu)\cos(\omega\hbar(\nu-i\beta)) }
                                               {1-\cosh(\omega\hbar\beta)}=\frac{1}{2} 
-\frac{1}{4}\frac{\sin\omega\hbar\nu \sin\omega\hbar(\nu-i\beta)}{\sinh^2(\frac{\omega\hbar\beta}{2})}
\nonumber \\
\frac{z'-y'}{\Delta}   &=& \frac{1}{2}\frac{\sin(\omega\hbar \nu)-\sin(\omega\hbar(\nu-i\beta))}
                                            {1-\cosh(\omega\hbar\beta)}  = -\frac{i}{2}\cos(\omega\hbar \nu)\coth(\frac{\omega\hbar\beta}{2}) 
+ \frac{1}{2}\sin(\omega\hbar \nu)
\nonumber
\end{eqnarray}
The functional ${\cal J}^{(1)}$ here vanishes completely, while the functionals ${\cal J}^{(2)}$
and  ${\cal J}^{(3)}$ vanish at $\nu=0$. 
Their derivatives at $\nu=0$ are
\begin{eqnarray}
\label{eq:J2-der-G}
\frac{\partial {\cal J}^{(2)}}{\partial\nu}|_{\nu=0} &=& -\frac{1}{4m}\int^t\int^t (XY'-X'Y)CC' \sin\omega(s-s') \coth(\frac{\omega\hbar\beta}{2}) \\
\label{eq:J3-der-G}
\frac{\partial {\cal J}^{(3)}}{\partial\nu}|_{\nu=0} &=& \frac{i}{4m}\int^t\int^t (XY'+X'Y)CC' \cos\omega(s-s') 
\end{eqnarray}
which are the  the same as 
${\cal I}^{(2)}$ and ${\cal I}^{(3)}$ in~\cite{Aurell2017},
with the idenification $\nu=i\epsilon$ and a factor two from the definition of the double integrals.
The second derivatives at $\nu=0$, which determine the variance of the change in bath energy, are
\begin{eqnarray}
\label{eq:J2-der2-G}
\frac{\partial^2 {\cal J}^{(2)}}{\partial\nu^2}|_{\nu=0} &=& -\frac{i\omega\hbar}{4m}\int^t\int^t (XY'-X'Y)CC' \sin\omega(s-s') \left(1+\coth^2(\frac{\omega\hbar\beta}{2})\right) \\
\label{eq:J3-der2-G}
\frac{\partial^2 {\cal J}^{(3)}}{\partial\nu^2}|_{\nu=0} &=& \frac{\omega\hbar}{4m}\int^t\int^t (XY'+X'Y)CC' \cos\omega(s-s')  \coth(\frac{\omega\hbar\beta}{2}) 
\end{eqnarray}

\end{widetext}

\bibliography{fluctuations}%
\end{document}